\documentclass[prb,amssymb,twocolumn,showpacs,superscriptaddress,floatfix]{revtex4}

\usepackage{graphicx}
\usepackage{bm}
\usepackage{mathptmx}

\begin{document}

\title{Critical behavior of repulsive linear $k$-mers on
triangular lattices}

\author{P. M. Pasinetti}
\affiliation{Departamento de F\'{\i}sica, Laboratorio de Ciencias
de Superficies y Medios Porosos, Universidad Nacional de San Luis,
CONICET, Chacabuco 917, 5700 San Luis, Argentina}
\author{F. Rom\'a}
\affiliation{Departamento de F\'{\i}sica, Laboratorio de Ciencias
de Superficies y Medios Porosos, Universidad Nacional de San Luis,
CONICET, Chacabuco 917, 5700 San Luis, Argentina}
\affiliation{Centro At\'omico Bariloche, Av. Bustillo 9500, 8400
S. C. de Bariloche, R\'{\i}o Negro, Argentina}
\author{J. L. Riccardo}
\affiliation{Departamento de F\'{\i}sica, Laboratorio de Ciencias
de Superficies y Medios Porosos, Universidad Nacional de San Luis,
CONICET, Chacabuco 917, 5700 San Luis, Argentina}
\author{A. J. Ramirez-Pastor}
\affiliation{Departamento de F\'{\i}sica, Laboratorio de Ciencias
de Superficies y Medios Porosos, Universidad Nacional de San Luis,
CONICET, Chacabuco 917, 5700 San Luis, Argentina}

\begin{abstract}
Monte Carlo (MC) simulations and finite-size scaling analysis have
been carried out to study the critical behavior in a submonolayer
two-dimensional gas of repulsive linear $k$-mers on a triangular
lattice at coverage $k/(2k+1)$. A low-temperature ordered phase,
characterized by a repetition of alternating files of adsorbed
$k$-mers separated by $k+1$ adjacent empty sites, is separated
from the disordered state by a order-disorder phase transition
occurring at a finite critical temperature, $T_c$. The MC
technique was combined with the recently reported Free Energy
Minimization Criterion Approach (FEMCA), [F. Rom\'a et al., Phys.
Rev. B, 68, 205407, (2003)], to predict the dependence of the
critical temperature of the order-disorder transformation. The
dependence on $k$ of the transition temperature, $T_c(k)$,
observed in MC is in qualitative agreement with FEMCA. In
addition, an accurate determination of the critical exponents has
been obtained for adsorbate sizes ranging between $k=1$ and $k=3$.
For $k>1$, the results reveal that the system does not belong to
the universality class of the two-dimensional Potts model with
$q=3$ ($k=1$, monomers). Based on symmetry concepts, we suggested
that the behavior observed for $k=1, 2$ and $3$ could be
generalized to include larger particle sizes ($k \geq 2$).

\end{abstract}

\date{\today}
\maketitle
\section{Introduction}
Two previous articles \cite{PRB4,PRB5}, referred to as Papers I
and II, respectively, were devoted to the study of repulsively
interacting $k$-mers on square lattices at half coverage. In Paper
I, Monte Carlo (MC) simulations were used to deal with such
problem. A $(2k \times 2)$ ordered phase, characterized by a
repetition of alternating files of adsorbed $k$-mers separated by
$k$ adjacent empty sites, was found. This ordered phase is
separated from the disordered state by a order-disorder phase
transition occurring at a critical temperature $T_c$. The $T_c$
dependence on the particle size shown an intriguing behavior, with
a pronounced minimum at $k=2$. In addition, two theoretical
approaches, Detailed Mean-Field Approximation (DMFA) and Free
Energy Minimization Criterion Approach (FEMCA),  were combined
with Monte Carlo simulations to obtain the critical temperature of
the order-disorder transformation. Predictions from FEMCA, based
on a free energy minimization criterion, shown a remarkable
qualitative agreement with the simulation data and allowed us to
interpret the physical meaning of the mechanisms underlying the
observed transitions. Paper II was a step further, analyzing the
universality class of the phase transition. For this purpose, an
extensive work of MC simulations and finite-size scaling analysis
was carried out. Based on the strong axial anisotropy of the
low-temperature phase for $k \geq 2$, a new order parameter
measuring the orientation of the admolecules was introduced.
Taking advantage of its definition, an accurate determination of
the critical exponents was obtained for adsorbate sizes ranging
between $k=2$ and $k=4$. The evaluation of the complete set of
static critical exponents, $\alpha$, $\beta$, $\gamma$ and $\nu$
for different molecular sizes, shown that: for linear $k$-mers
with $k \geq 2$, this phase transition does not belong to the
universality class of the two-dimensional Ising model ($k=1$). The
main source for this behavior was associated to the breaking of
the orientational symmetry occurring for $k \geq 2$, which does
not occur for $k=1$. Moreover, the critical exponents reported in
Paper II do not correspond to a known universality class,
according to the current classification of order-disorder
transitions on surfaces given by M. Schick \cite{Schick}. The aim
of this paper is the study of repulsive straight $k$-mers on
triangular lattices using the same techniques developed in Papers
I and II.

The behavior of interacting dimers has also been analyzed by using
graph theory \cite{Seitz}. In Ref. \cite{Seitz}, a Sierpinski
gasket with a fractal dimension of $\ln 3 / \ln 2$ was used as
substrate. However, the technique can be extend to other fractal
graphs with low ``ramification degree". With respect to triangular
lattices, leading contributions have been published by Hock and
McQuistan \cite{Hock} and Phares and Wunderlich \cite{Phares}. The
thermodynamics of non-interacting dimers was obtained from the
knowledge of the occupational degeneracy of dimers partially
covering the lattice. Later, R\.zysko and Bor\'owko \cite{Rzysko}
studied the interesting problem of heteronuclear dimers consisting
of different segments $A$ and $B$ adsorbed on square and
triangular lattices. The authors considered models with attractive
$B-B$ and $A-B$ nearest-neighbor energies and variable $A-A$
energy (attractive as well as repulsive). A rich variety of phase
transitions was reported along with a detailed discussion about
critical exponents and universality class. In the case of a
triangular lattice all systems belong to the class of universality
of the two-dimensional Potts model with $q=3$.

In previous work we have studied the adsorption thermodynamics of
attractive and repulsive dimers on honeycomb and triangular
lattices \cite{LANG7}. From the study of adsorption properties
such as adsorption isotherm, heat of adsorption and
configurational entropy, two different ordered phases occurring in
the adlayer were reported for repulsive couplings and low
temperatures. Namely, $(1)$ the low coverage ordered phase (LCOP),
with $2/5$ ($5/9$) of the sites occupied for a triangular
(honeycomb) lattice and $(2)$ the high coverage ordered phase
(HCOP), with $2/3$ of the sites filled for both lattices. In the
particular case of triangular lattices, the LCOP resembles the $(4
\times 2)$ ordered structure of repulsive dimers on square
lattices at half coverage. As in the last case, from the LCOP
appearing in dimers at $\theta=2/5$, we can predict the existence
of a ordered structure for repulsive straight $k$-mers on
triangular lattices at coverage $k/(2k+1)$. A snapshot of this
low-temperature phase, characterized by a repetition of
alternating files of adsorbed $k$-mers separated by $k+1$ adjacent
empty sites, is shown in Fig. 1 for $k=2$.

In this context, the scope of the present work is: $(i)$ to calculate,
via MC simulations \cite{Binder,Nicholson,MC}, finite-size scaling analysis \cite{Fss},
and FEMCA, the critical temperature as a function of the size of
the adsorbed molecules for repulsive linear $k$-mers adsorbed on
triangular lattices at $\theta=k/(2k+1)$;
and $(ii)$ to determine critical exponents and universality.
The outline of the paper is as follows: In Section 2 we describe the lattice-gas
model and the simulation scheme. In Section 3 we present the results.
Finally, the general conclusions are given in Section 4.

\section{Lattice-gas model and Monte Carlo simulation scheme}
\subsection{The model}

In this section we describe the lattice-gas model for the
adsorption of particles with multisite occupancy in the monolayer
regime. We consider the adsorption of homonuclear $k$-mers on
triangular lattices. The adsorbate molecules are assumed to be
composed by $k$ identical units in a linear array with constant
bond length equal to the lattice constant $a$. The $k$-mers can
only adsorb flat on the surface occupying $k$ lattice sites. The
surface is represented as a triangular array of $M=L \times L$
adsorptive sites, where $L$ is the linear size of the array. In
order to describe a system of $N$ $k$-mers adsorbed on $M$ sites
at a given temperature $T$, let us introduce the occupation
variable $c_i$ which can take the values $c_i=0$ or $1$, if the
site $i$ is empty or occupied by a $k$-mer unit, respectively. The
$k$-mer retain its structure upon adsorption, desorption and
diffusion. The Hamiltonian of the system is given by,
\begin{equation}
H = w \sum_{\langle i,j \rangle} c_i c_j - N(k-1) w+ \epsilon_o \sum_{i} c_i \label{h}
\end{equation}
where $w$ is the nearest-neighbor (NN) interaction constant which
is assumed to be repulsive (positive),$\langle i,j \rangle$
represents pairs of NN sites and $\epsilon_o$ is the energy of
adsorption of one given surface site. The term $N(k-1)w$ is
subtracted in Eq. (\ref{h})  since the summation over all the
pairs of NN sites overestimates the total energy by including
$N(k-1)$ bonds belonging to the $N$ adsorbed $k$-mers. Finally,
$\epsilon_o$ is set equal zero, without any lost of generality.

\subsection{Order parameter and Monte Carlo method}

In order to study the order-disorder phase transition occurring in
the adsorbate, it is convenient to define a related order
parameter. For $k=1$, a standard geometrical order parameter can
be built \cite{Landau}. For this purpose, the lattice is separated
in three intersecting sublattices, $\rho = 1, 2, 3$, and for
each one of them the functions:
\begin{equation}
f_{\rho}=\frac{3}{M} \sum_{i \in \rho} c_i
\end{equation}
are obtained. Then the order parameter $\varphi$ of the system is
given by:
\begin{equation}
\varphi=\frac{4}{\sqrt{6}} \left( \sum_{\rho} \varphi_{\rho}^2
\right)^{1/2}
\end{equation}
where
\begin{equation}
\varphi_{\rho}=\frac{\delta_{\rho \tau \chi}}{2} \left[
f_{\rho} - \frac{\left( f_{\tau} + f_{\chi} \right)}{2}
\right]
\end{equation}
and the $\delta$ takes the value $1$ for a cyclic permutation of
subindexes $(1,2,3)$ and the value $0$ otherwise.

As in Paper II, an order parameter measuring the orientation of
the admolecules in the ordered structure will be defined for
$k>1$. Fig. 1 shows a snapshot corresponding to one of the
possible configurations of the low-temperature phase for dimers at
$\theta=2/5$. As it can be observed, the structure adopts an
orientation along one of the lattice's axis. Taking advantage of
this property, we define $\delta_k$ as:
\begin{equation}
\delta_k =  \frac{1}{2} \left( \left | \frac{N_1-N_2}{N}  \right |
+ \left | \frac{N_2-N_3}{N}  \right | + \left | \frac{N_3-N_1}{N}
\right | \right) \label{fi2}
\end{equation}
where $N_x$ ($x=1,2,3$) represents the number of $k$-mers aligned
along one of the three axis of the lattice and $N=N_1+N_2+N_3$. It
is worth to notice that, due to the low-temperature phase is
isotropic for monomers, $\delta_k$ is restricted to $k\geq 2$.

When the system is disordered $(T>T_c)$, the three orientations
 are equivalents and $\delta_k$ is zero. As the temperature is decreased
below $T_c$, the $k$-mers align along one direction and $\delta_k$
is different from zero (being $\delta_k=1$ at $T=0$). Then,
$\delta_k$ appears as a good order parameter, evidencing the
order-disorder phase transition.

As it is well-known for $k=1$, the system belongs to the
universality class of the three-state Potts model
\cite{Schick,Baxter,Landau,Yeomans}, being $\nu=5/6$, $\beta=1/9$,
$\gamma=13/9$ and $\alpha=1/3$. In this case, the phase transition
is accomplished by a breaking of the translational symmetry (the
coverage of the sublattices is different for $T<T_c$). However, an
additional breaking of the orientational symmetry occurs for $k
\geq 2$. Consequently, a change in the universality class is
expected for linear molecules with $k \geq 2$, with respect to the
case of monomers.

In order to study the critical behavior of the system, we have
used an efficient exchange MC method \cite{Hukushima,Earl} and
finite-size scaling analysis \cite{Binder,Fss,Fisher}. As in Ref.
\cite{Hukushima}, we build a compound system which consists of $m$
non-interacting replicas of the system concerned. The $m$-th
replica is associated with the temperature $T_m$ [or
$\beta_m=1/(k_B T_m)$, being $k_B$ the Boltzmann constant]. In
other words, each replica is in contact with its own heat bath
having different temperature. Under these conditions, the
algorithm to carry out the simulation process is the following:

\begin{itemize}
\item[1)] The compound system of $m$ replicas is generated. For
this purpose, each replica is simulated simultaneously and
independently as canonical ensemble for $n_1$ MC steps by using a
standard importance sampling MC method
\cite{Binder,Nicholson,Metropolis}. In order to determine the set
of temperatures, $\{\beta_m \}$, we set the highest temperature,
$T_{max} (\beta_{max})$, in the high temperature phase where
relaxation (correlation) time is expected to be very short and
there exists only one minimum in the free energy space. On the
other hand, the lowest temperature, $T_{min} (\beta_{min})$, is
somewhere in the low-temperature phase whose properties we are
interested in. Finally, the difference between two consecutive
temperatures is set as $\left(T_{max} - T_{min} \right)/(m-1)$
(equally spaced temperatures).

\item[2)] Interchange vacancy-particle and diffusional relaxation
\cite{foot1}. The procedure is as follows:
\begin{itemize}
\item[2.1)] One of the $m$ replicas is randomly selected;
\item[2.2)] a $k$-mer and a linear $k$-uple of empty sites, both
belonging to the replica chosen in 2.1), are randomly selected and
their positions are established. Then, an attempt is made to
interchange its occupancy state with probability given by the
Metropolis rule \cite{Metropolis}:
\begin{equation}
P = \min \left\{1,\exp\left( - \beta_m \Delta H \right) \right\}
\end{equation}
where $\Delta H=H_f -H_i$ is the difference between the
Hamiltonians of the final and initial states;

\item[2.3)] a $k$-mer is randomly selected. Then, a displacement
to nearest neighbor positions is attempted (following the
Metropolis scheme), by either jumps along the $k$-mer axis or
reptation by rotation around a unity of the $k$-mer. This
procedure (diffusional relaxation) must be allowed in order to
reach equilibrium in a reasonable time.

\end{itemize}

\item[3)] Exchange of two configurations $X_m$ and $X_{m'}$,
corresponding to the $m$-th and $m'$-th replicas, respectively,
is tried and accepted with the probability $W\left(X_m,\beta_m|
X_{m'},\beta_{m'}\right)$. In general, the probability of
exchanging configurations of the $m$-th and $m'$-th replicas is
given by \cite{Hukushima},
\begin{equation}
W\left(X_m,\beta_m| X_{m'},\beta_{m'}\right)=\left\{
\begin{array}{cc}
1 & {\rm for}\ \ {\Delta<0} \\
\exp(-\Delta)  & {\rm for}\ \ {\Delta>0}
\end{array}
\right.
\end{equation}
where $\Delta=\left( \beta_m - \beta_{m'} \right)\left[ H(X_m') -
H(X_{m}) \right]$. As in Ref. \cite{Hukushima}, we restrict the
replica-exchange to the case $m \leftrightarrow m +1$.

\item[4)] Repeat from step 2) $m \times M $ times. This is the
elementary step in the simulation process or Monte Carlo step
(MCS).


\end{itemize}

The procedure 1)-4) is repeated for all lattice's sizes. For each
size, the equilibrium state can be well reproduced after
discarding the first $n_2$ MCS. Then, averages are taken over
$n_{MCS}$ successive MCS. The canonical expectation value of a
physical quantity $A$ is obtained in the usual way as follows:
\begin{equation}
{\langle A\rangle}_{\beta_m}=\frac{1}{n_{MCS}}
\sum_{t=1}^{n_{MCS}} A \left[X_m(t)\right]
\end{equation}

All calculations were carried out using the parallel cluster BACO
of Universidad Nacional de San Luis, Argentina. This facility
consists of 60 PCs each with a 3.0 MHz Pentium-4 processors. The
values of the parameters used in the simulated tempering runs are
shown in Table I.

The internal energy per lattice site, $u$, is obtained as simple
averages:

\begin{equation}
u=\frac{1}{L^2} \left< H \right>_T.
\end{equation}

 The specific heat $C$ is sampled from energy fluctuations,

\begin{equation}
C= \frac{1}{L^2 k_BT^2} [\langle H^2 \rangle_T - \langle H \rangle
^2_T].
\end{equation}

The quantities related with the order parameter, such as the
susceptibility $\chi$, and the reduced fourth-order cumulant $U_L$
introduced by Binder \cite{Binder}, can be calculated as:

\begin{equation}
\chi = \frac{L^2}{k_BT} [ \langle z^2 \rangle_T - \langle z
\rangle^2_T]
\end{equation}

\begin{equation}
U_L(T) = 1 -\frac{\langle z^4\rangle_T} {3\langle z^2\rangle_T^2}
\label{cum}
\end{equation}

\noindent where $z$($\equiv \varphi$ or $\delta_k$) represents the
order parameter and the thermal average $\langle ... \rangle_T$,
in all the quantities, means the time average throughout the MC
simulation.

\section{Results}

The calculations were developed for linear $k$-mers, with $k$
ranging between $1$ and $3$, on a triangular $L\times L$ lattice
at coverage $k/(2k+1)$. In addition, conventional periodic
boundary conditions were considered. Note, however, that the
choice of appropriate linear dimensions $L$ has to be done in such
away that the ordered structures are not disturbed. Thus, for
$k=1$, the effect of finite size was investigated by examining
lattices varying from $L=30$ to $L=120$. In the case of $k=2$ and
$k=3$, $L= 10, 15, 20, 25, 30$ and $L= 14, 21, 28, 35$ were used,
respectively, with an effort reaching almost the limits of our
computational capabilities.


The critical temperatures have been estimated from the plots of
the reduced four-order cumulants $U_L(T)$ plotted versus  $k_BT/w$
for several lattice sizes. In the vicinity of the critical point,
cumulants show a strong dependence on the system size. However, at
the critical point the cumulants adopt a nontrivial value $U^*$;
irrespective of system sizes in the scaling limit. Thus, plotting
$U_L(T)$ for different linear dimensions yields an intersection
point $U^*$, which gives an accurate estimation of the critical
temperature in the infinite system.

Hereafter we discuss the behavior of the critical temperature as a
function of the size $k$, for adsorbed $k$-mers at monolayer. Fig.
2 illustrates the reduced four-order cumulants $U_L(T)$ plotted
versus  $k_BT/w$ for several lattice sizes and $k=1$. From their
intersections one gets the estimation of the critical temperature.
The value obtained for the critical temperature was $k_BT_c/w =
0.3354(1)$, being $U^* = 0.617(1)$. Our calculation of the
critical temperature is in well agreement with the transfer-matrix
result of Kinzel and Schick \cite{Kinzel}. With respect to Monte
Carlo calculations, Metcalf \cite{Metcalf} in the 70's and Chin
and Landau \cite{Chin} in the 80's obtained similar values of the
critical temperature. Namely, $k_BT_c/w \approx 0.35$. Due to the
lattice sizes and the number of MCS used in this contribution, our
estimate of $T_c$ is expected to be more accurate than those
reported previously.

Figs. 3 and 4 show the results of $U_L(T)$ for $k = 2$ and $k =
3$, respectively. In the case $k=2$, the value estimated for
$k_BT_c/w$ [$=0.2338(1)$] agrees very well with previous
determinations reported in the literature \cite{LANG7}. In Ref.
\cite{LANG7}, a value $k_BT_c/w \approx 0.23$ was obtained from
the inflection on the function $s(T)$, being $s(T)$ the
configurational entropy of the adlayer as a function of the
temperature. With respect to $k=3$, the value obtained for
$k_BT_c/w$ [$=0.341(1)$] is reported for the first time in the
literature.

The values of the critical temperature, which are collected in
Table II for $k=1-3$, present a non-trivial behavior as a function
of the particle size $k$. To understand the dependence of $T_c(k)$
on $k$, it is convenient to appeal to the recently reported FEMCA
\cite{PRB4}. In this theoretical framework, $T_c$ depends on the
ratio of the energy per lattice site and entropy per lattice site
differences, $u_{\infty}-u_0$ and $s_{\infty}-s_0$, respectively,
between a fully disordered state ($T \rightarrow \infty$) and the
ground state ($T \rightarrow 0$),
\begin{equation}
T_c  \approx \frac{\Delta u}{\Delta s} = \frac{u_{\infty}-u_0}{s_{\infty}-s_0} \label{fetc1}
\end{equation}
In this case, the mean energy and the entropy for the ordered
state are $u_0 = s_0 = 0$. Then, the critical temperature depends
on the mean energy and the entropy of the disordered state.
$u_{\infty}$ can be calculated from the quasi-chemical
approximation \cite{SURFSCI12},
\begin{equation}
\frac{u_{\infty}}{ w} = \left( \frac{\lambda \theta}{2k}- \alpha
\right) \stackrel{c=6, T \rightarrow \infty; \atop \theta=
k/(2k+1) }{=} \frac{2k+1}{ 5k+4}, \label{u2}
\end{equation}
being $c$ the lattice connectivity. In addition,
\begin{equation}
\lambda=(c-2)k+2,
\end{equation}
\begin{equation}
\alpha = \frac{\lambda c}{2k} \frac{\theta
(1-\theta)}{\left[\frac{c}{2}-\left(\frac{k-1}{k} \right)\theta +
b\right]},    \label{alfa}
\end{equation}
and
\begin{equation}
b= \left\{ \left[\frac{c}{2}-\left(\frac{k-1}{k} \right)\theta
\right]^2  - \frac{\lambda c}{k} \left[ 1-\exp(-w/k_BT) \right]
\theta (1-\theta) \right\}^{1/2}.    \label{b}
\end{equation}

With respect to the entropy,

\begin{equation}
s_{\infty}/k_B=\ln 3 - (2/3)\ln 2    \ \ \ \ \ \ \ \ \ \     {\rm
for} \ \ k=1, \label{sk1}
\end{equation}

and

\begin{eqnarray}
\frac{s_{\infty}}{k_B} & = & \frac{\theta}{k} \ln \frac{c}{2}
-(1-\theta)\ln(1-\theta) - \frac{\theta}{k} \ln \frac{\theta}{k} \nonumber\\
& & + \left[ 1 - \frac{(k-1)}{k} \theta   \right]  \ln \left[ 1-  \frac{(k-1)}{k} \theta  \right]  \nonumber\\
& \stackrel{c=6; \atop \theta= k/(2k+1)}{=} &  \frac{\ln 3}{2k+1} -\frac{k+1}{2k+1}\ln \left( \frac{k+1}{2k+1} \right) \nonumber\\
& & -\frac{1}{2k+1} \ln \left(\frac{1}{2k+1} \right) \nonumber\\
& & + \frac{k+2}{2k+1} \ln \left( \frac{k+2}{2k+1} \right) \ \ \ \
\ \ \ {\rm for}\ \ k \geq 2.  \label{s}
\end{eqnarray}

In the last equation, the entropy was obtained from the extension
to higher dimensions of the exact configurational factor of linear
chains adsorbed on one-dimensional lattices \cite{LANG9}. Under
these considerations, the critical temperature results,
\begin{eqnarray}
& &\frac{k_B T_c(k)}{w}  \approx  \frac{u_{\infty}/w}{s_{\infty}/k_B} \label{tcfinal} \\
& & \approx  \left\{
\begin{array}{cc}
1/\left( 3 \ln3 - 2 \ln 2 \right)= 0.5237 & {\rm for}\ \ k=1 \\
\frac{(2k+1)^2}{\left\{(5k+4) \left[\ln 3 -(k+1)\ln \left(
\frac{k+1}{2k+1} \right) -\ln \left( \frac{1}{2k+1} \right) +
\left( k + 2 \right) \ln \left( \frac{k+2}{2k+1}
\right)\right]\right\}} & {\rm for}\ \ k \geq 2
\end{array} \right. \nonumber
\end{eqnarray}

Fig. 5 presents the theoretical predictions obtained from FEMCA
(Eq. \ref{tcfinal}) for the critical temperature as a function of
the size $k$. For high values of $k$, $k_B T_c(k)/w$ increases
monotonically as $k$ is increased (see inset). In the other
extreme, the theoretical curve shows a marked break at $k=2$,
which is in good qualitative agreement with the minimum obtained
from Monte Carlo simulations.

The theoretical approach allows us to interpret the physical
meaning of the main features of the critical temperature. In fact,
Eq. (\ref{fetc1}) shows that $k_B T_c/w$ depends on the mean
energy and the entropy of the disordered state. In first term, we
will analyze the entropy. When one pass from $(k=1,\theta=1/3)$ to
$(k=2,\theta=2/5)$, the number of entities per lattice site,
$n_e$, and for this reason, the number of accessible states,
diminishes [in general, $n_e(k) \propto 1/(2k+1)$]. On the other
hand, a new degree of freedom appears for the adsorbed particles
whose size is $k \geq 2$: dimers, trimers, etc. can rotate on the
lattice. The new accessible states for dimers due to possible
rotations compensate the diminution in the number of entities with
respect to monomers. Consequently, the variation in the entropy
between monomers and dimers is small. Thus,
$s_{\infty}(k=1,\theta=1/3)=0.6365$ [see Eq. (\ref{sk1})] and
$s_{\infty}(k=2,\theta=2/5)\approx 0.6339$ \cite{LANG7} for
monomers and dimers adsorbed in triangular lattices, respectively.
However, due to the new degree of freedom appears for $k \geq 2$,
the diminution in the entropy is reestablished (associated to the
diminution in the number of entities) and $s_{\infty}$ diminishes
as $k$ is increased for $k>2$.

With respect to the mean energy, the approximated solution given
in Eq. (\ref{u2}) shows that $u_{\infty}(k)$ increases until an
asymptotic value for higher $k$'s: $u_{\infty}(1)=0.333,
u_{\infty}(2)=0.357, u_{\infty}(3)=0.368, u_{\infty}(4)=0.375,
\cdots, u_{\infty}(\infty)=0.40$.

Now we can interpret the two different regimes observed in the
curve of Fig. 5. While $u_{\infty}(k)$ changes slightly in all
range of $k$, $s_{\infty}(k)$ presents an abrupt change at $k=2$.
Physically, this change in the entropy, which is responsible for
the marked break appearing in the curve of $k_B T_c(k)/w$ versus
$k$, is associated with the possibility of orientation of the
adsorbed molecules for $k \geq 2$.

As it was shown in Figs. 3 and 4, the intersections of the
cumulants for $k=2$ and $k=3$ suggest that the system suffers a
phase transition at coverage $k/(2k+1)$. In order to corroborate
this finding, the critical exponents will be calculated in the
following.


In first term, we study exhaustively the well-known lattice-gas of
repulsive monomers. Once we know $T_c$ (see Fig. 2), the critical
exponent $\nu$ can be calculated from the full data collapsing of
$U_L$. The results are shown in Fig. 6 (being $t \equiv T/T_c
-1)$, where an excellent fit was obtained for $\nu =5/6$. Given
$k_B T_c / w=0.3354(1)$ and $\nu=5/6$, $\alpha$, $\beta$ and
$\gamma$ were obtained from the collapse of the curves of $C$,
$\varphi$ and $\chi$, as it is shown in Figs. 7-9, respectively.
The data scaled extremely well using $\alpha=1/3$, $\beta=1/9$ and
$\gamma=13/9$, which corroborates that the case of $k=1$
corresponds to the universality class of the two-dimensional Potts
model with $q=3$. The values of the critical exponents along with
the intersection point of the cumulants are collected in Table II.

The finite-size scaling study was carried out for $k=2$ (Figs.
10-13) and $k=3$ (data do not shown here). The resulting values of
the critical exponents, which are listed in Table II, point out
clearly the existence of a different universality class from that
of the three-state Potts model.

As can be demonstrated, the set of critical exponents for dimers
and trimers fulfills the well-known inequalities of Rushbrooke
\cite{Rushbrooke}, $\alpha + 2\beta+\gamma \geq 2$, and Josephson
\cite{Josephson}, $d \nu + \alpha \geq 2$ (being $d$ the dimension
of the space), leading to independent controls and consistency
checks of the values of all the critical exponents.

The identical results (within numerical errors) obtained for the
critical exponents corresponding to dimers and trimers corroborate
our hypothesis that the breaking of the orientational symmetry
occurring for $k \geq 2$: $i)$ affects the nature of the phase
transition and $ii)$ is the main source of the change in the
universality class, with respect to the well-known two-dimensional
three-state Potts class of universality corresponding to monomers.

Surprisingly, the exponents obtained for repulsive $k$-mers
($k>1$) on a triangular lattice at $\theta=k/(2k+1)$ are very
similar to those characterizing the critical behavior of $k$-mers
($k>1$) on square lattices at $\theta=1/2$ \cite{PRB5}. However,
we think that these two phase transitions belong to different
universality classes. Some evidences support our arguments: $1)$
the value obtained for the fixed-point cumulant $U^*$ varies from
square to triangular lattices \cite{foot2}; and $2)$ the
low-temperature phase  have two and three possible orientations
for square and triangular lattices, respectively. Then, the order
parameters characterizing both phase transitions have different
symmetries.

\section{Conclusions}

In the present work, we have addressed the critical properties of repulsive linear $k$-mers
on two-dimensional triangular lattices at coverage $k/(2k+1)$, and shown the dependence of the
critical temperature on the size $k$. The results were obtained by using MC simulations
and theoretical calculations from FEMCA.

The critical temperature dependence on the particle size of the
low coverage ordered phase [$\theta=k/(2k+1)$] of repulsive
straight particles have been reported for the first time, and
found that dimers present the minimum value. FEMCA shown that an
analysis of the delicate balance between the size dependence of
entropy and energy per site allows us to interpret qualitatively
the behavior of $T_{c}$ versus $k$.

On the other hand, the evaluation of the complete set of static
critical exponents, $\alpha$, $\beta$, $\gamma$ and $\nu$ for
different molecular sizes, shows that for linear $k$-mers with $k
\geq 2$, this phase transition does not belong to the universality
class of the two-dimensional Potts model with $q=3$. The main
source for this behavior is the breaking of the orientational
symmetry occurring for $k \geq 2$, which does not occur for $k=1$.
Moreover, the critical exponents reported in the present paper for
$k \geq 2$ do not correspond to a known universality class,
according to the current classification of order-disorder
transitions on surfaces given by Schick \cite{Schick}.

A detailed scaling analysis shows that usual hyperscaling
relations (inequalities of Rushbrooke and Josephson) are fulfilled
and leads to independent controls of the values of all
the critical exponents.

Future efforts will be directed to obtain
the critical behavior for other existing ordered phases in the whole
range of coverage.


\newpage

\section{Tables and Figure Captions}

\begin{itemize}

\item[Table I] Parameters of the simulated tempering runs.

\item[Table II] Values of $T_c$, $\alpha$, $\beta$, $\gamma$,
$\nu$ and $U^*$ (as indicated in the text) for $k$ ranging from
$1$ to $3$. In the case $k=1$, the critical exponents correspond
to the well-known two-dimensional Potts model with $q=3$.

\item[Figure 1:] Snapshot of the ordered phase for dimers at
$\theta=2/5$.

\item[Figure 2:] Curves of $U_L(T)$ versus  $k_BT/w$, for $k=1$
and different lattice sizes as indicated. From their intersections
one obtained $k_BT_c/w$.

\item[Figure 3:] Same as Fig. 2 for dimers ($k=2$).

\item[Figure 4:] Same as Fig. 2 for trimers ($k=3$).

\item[Figure 5:] Comparison between simulated and theoretical
results for $k_BT_c/w$ vs. $k$. The symbology is indicated in the
figure.

\item[Figure 6:] Data collapsing for the cumulants in Fig. 2.

\item[Figure 7:] a) Size dependence of the specific heat, $C$
(with $k=1$) as a function of temperature. b) Data collapsing ($C
L^{-\alpha/\nu}$ vs. $t L^{1/\nu}$) for the curves in a).

\item[Figure 8:] a) Size dependence of the order parameter,
$\varphi$ (with $k=1$) as a function of temperature. b) Data
collapsing ($\varphi L^{\beta/\nu}$ vs. $\left|t\right|
L^{1/\nu}$) for the curves in a).

\item[Figure 9:] a) Size dependence of the susceptibility, $\chi$
(with $k=1$) as a function of temperature. b) Data collapsing
($\chi L^{-\gamma/\nu}$ vs. $t L^{1/\nu}$) for the curves in a).

\item[Figure 10:] Data collapsing for the cumulants in Fig. 3.

\item[Figure 11:] Same as Fig. 7 for dimers ($k=2$).

\item[Figure 12:] Same as Fig. 8 for dimers ($k=2$).

\item[Figure 13:] Same as Fig. 9 for dimers ($k=2$).
\end{itemize}

\newpage

\begin{tabbing}
\end{tabbing}
\begin{center}
TABLE I
\end{center}
$$
\begin{array}{|c|c|c|c|c|c|c|c|}
\hline  \hline
k & L & m &  n_1 & n_2 & n_{MCS} & T_{min} & T_{max} \\
\hline
   & 30 & 40 &  10^5 &  10^5 & 5 \times 10^5 & 0.279 & 0.392 \\
\cline{2-8}
1  & 60 & 40 &  10^5 &  10^5 & 5 \times 10^5 & 0.311 & 0.360 \\
\cline{2-8}
  &  90 & 40 &  10^5 &  10^5 & 5 \times 10^5 & 0.290 & 0.381 \\
\cline{2-8}
  &  120 & 40 &  10^5 &  10^5 & 5 \times 10^5 & 0.325 & 0.346 \\
\hline \hline
   & 10 & 41 &  5 \times 10^4 &  5 \times 10^4 & 10^5 & 0.22 & 0.30 \\
\cline{2-8}
  & 15 & 41 &  5 \times 10^4 &  5 \times 10^4 & 10^5 & 0.22 & 0.27 \\
\cline{2-8}
2  &  20 & 61 &  5 \times 10^4 &  5 \times 10^4 & 10^5 & 0.22 & 0.26 \\
\cline{2-8}
  &  25 & 101 &  5 \times 10^4 &  5 \times 10^4 & 10^5 & 0.23 & 0.25 \\
\cline{2-8}
  &  30 & 131 &  2.5 \times 10^4 &  2.5 \times 10^4 & 5 \times 10^5 & 0.23 & 0.25 \\
\hline \hline
   & 14 & 101 &  1.5 \times 10^5 &  1.5 \times 10^5 & 3 \times 10^5 & 0.300 & 0.400 \\
\cline{2-8}
  & 21 & 61 &  2.5 \times 10^5  &  2.5 \times 10^5 & 5 \times 10^5 & 0.320 & 0.380 \\
\cline{2-8}
3  &  28 & 51 &   2.5 \times 10^5 & 2.5 \times 10^5 & 5 \times 10^5 & 0.335 & 0.360 \\
\cline{2-8}
  &  35 & 51  &   10^6 &    10^6 & 2 \times 10^6 & 0.337 & 0.347 \\
\hline

\end{array}
$$


\newpage

\begin{tabbing}
\end{tabbing}
\begin{center}
TABLE II
\end{center}
$$
\begin{array}{|c|c|c|c|c|c|c|}
\hline  \hline
k & k_BT_c/w & {\it \alpha} &  {\it \beta} & {\it \gamma} & {\it \nu} & U^* \\
\hline
1   & 0.3354(1) & 1/3 &  1/9 & 13/9 & 5/6 & 0.617(1) \\
\hline
2  & 0.2338(1) & 0.95(3) &  \approx 0.0005 & 1.17(3) & 0.50(1) & 0.656(1) \\
\hline
3  &  0.341(1) & 1.00(4) &  \approx 0.0000 & 1.20(4) & 0.50(1) & 0.656(2) \\
\hline
\end{array}
$$

\newpage

\begin{figure}
\includegraphics[angle=-90,width=12cm]{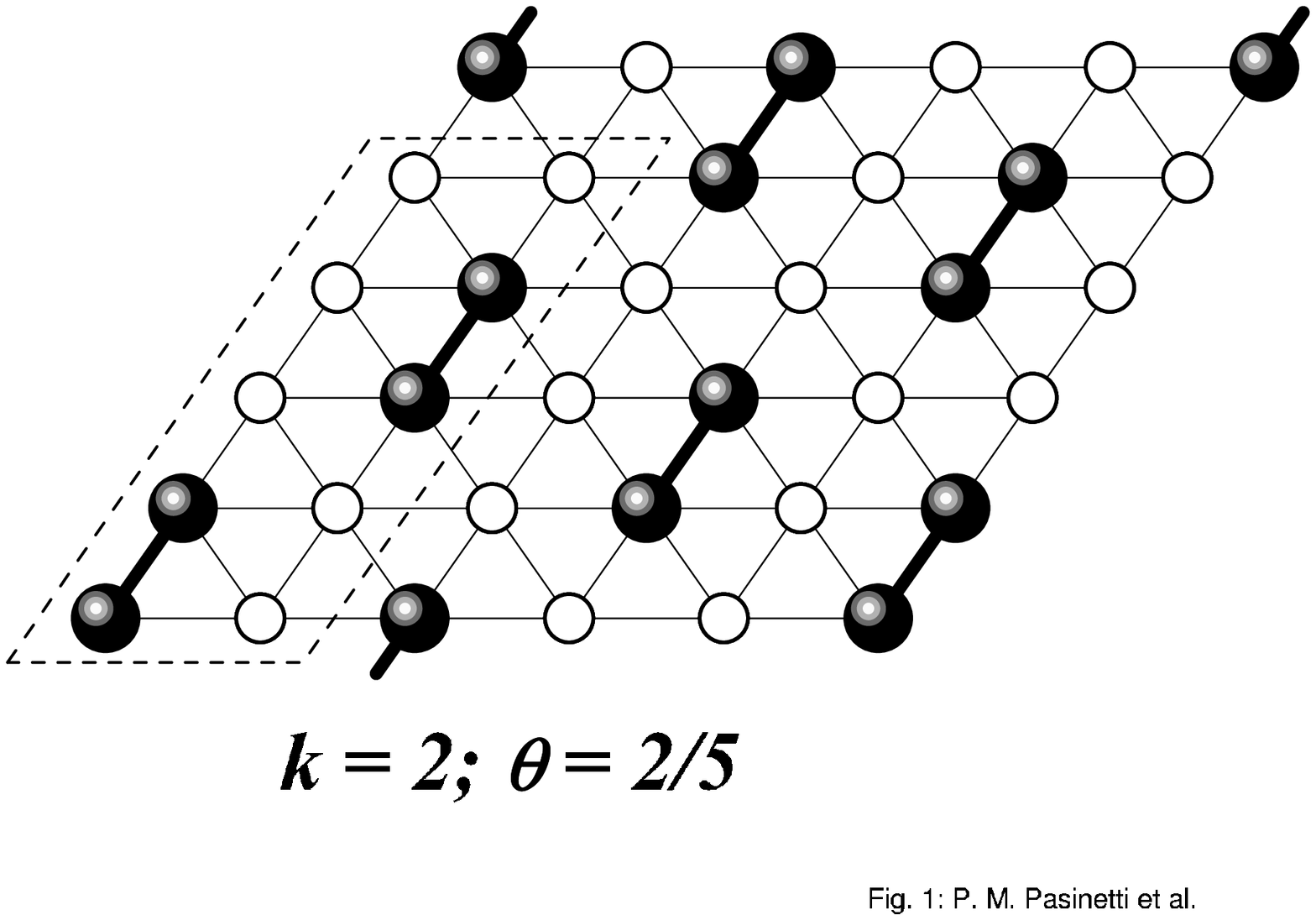}
\end{figure}
\begin{figure}
\includegraphics[angle=0,width=12cm]{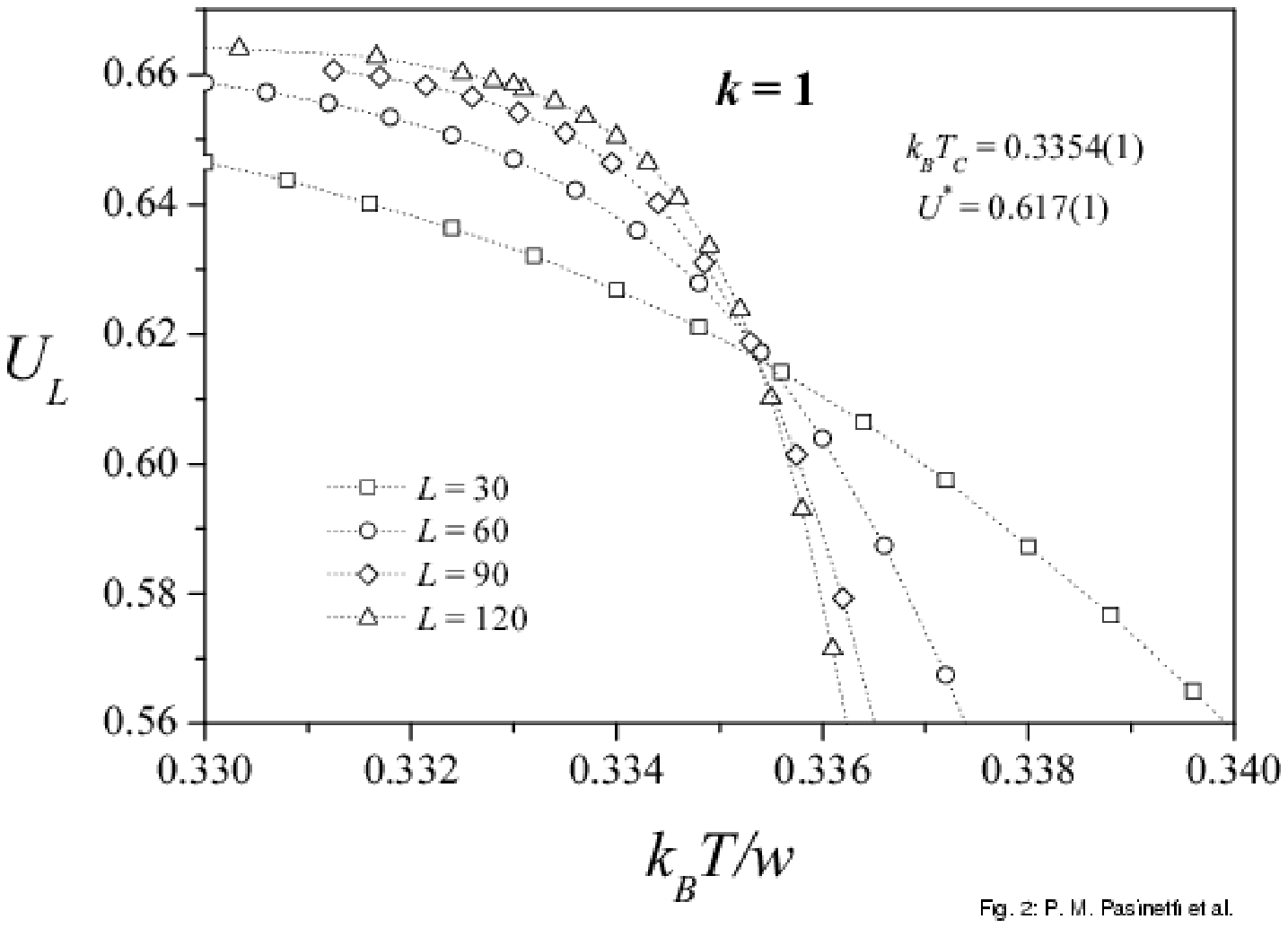}
\end{figure}
\begin{figure}
\includegraphics[angle=0,width=12cm]{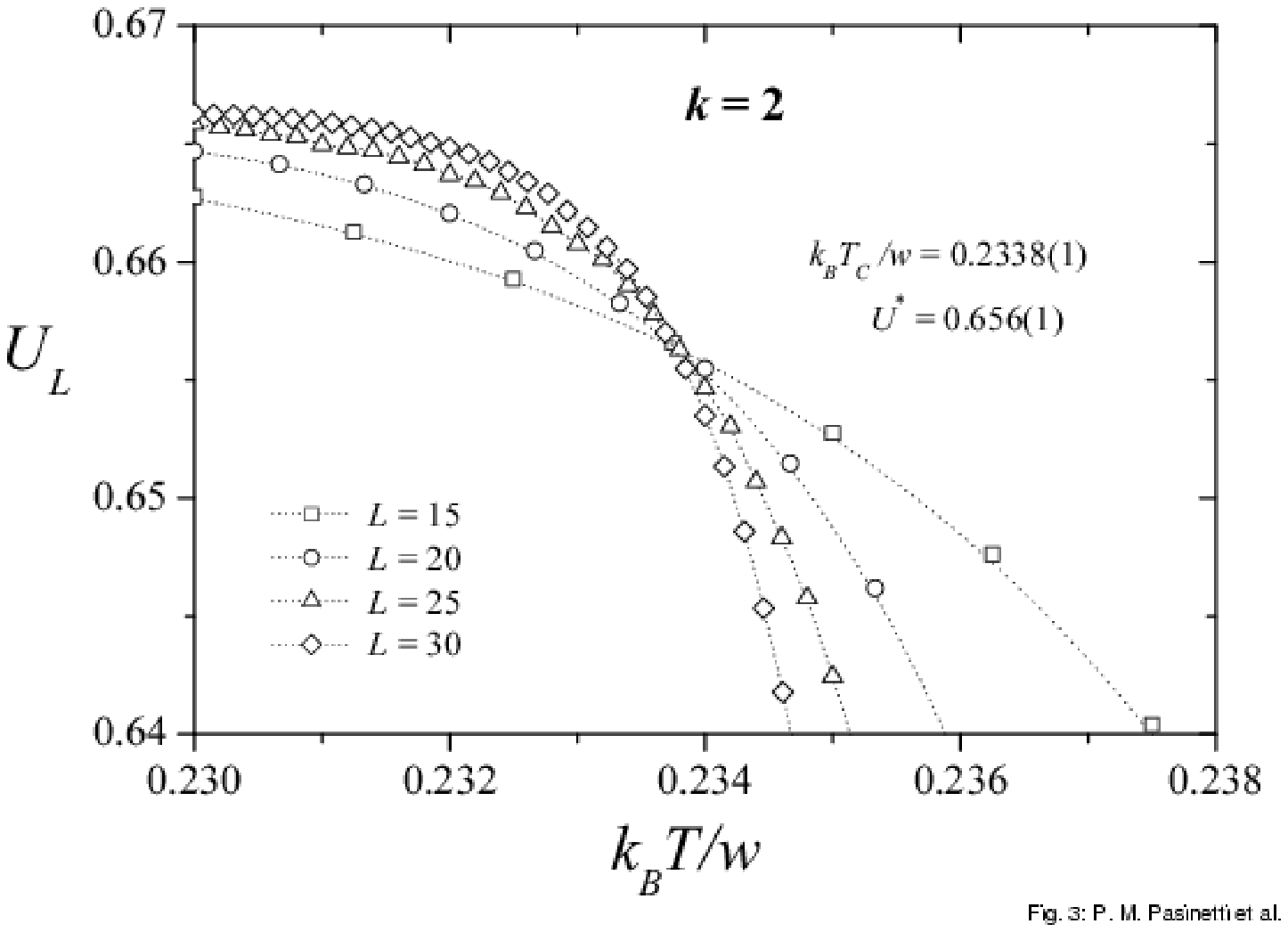}
\end{figure}
\begin{figure}
\includegraphics[angle=0,width=12cm]{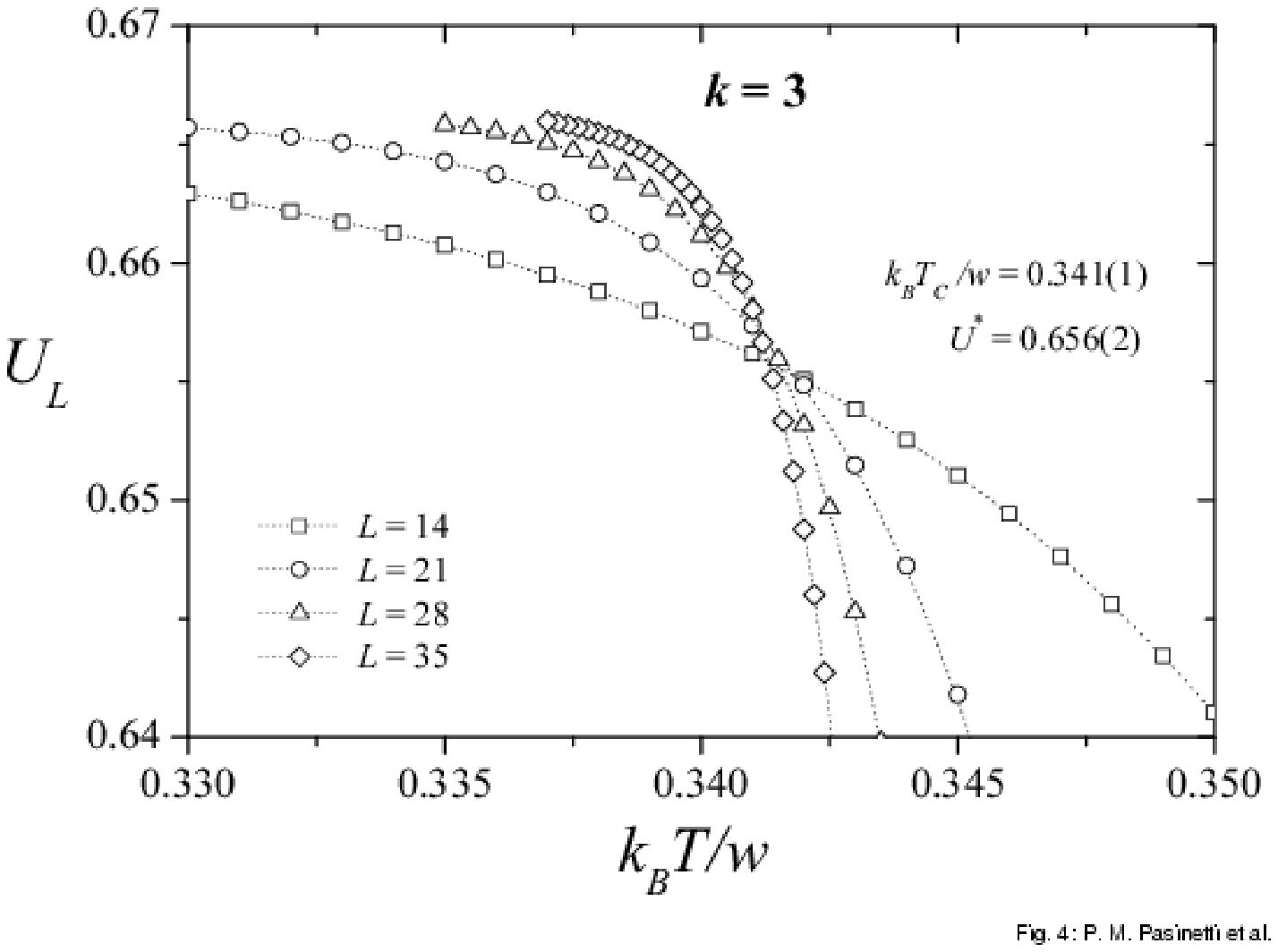}
\end{figure}
\begin{figure}
\includegraphics[angle=0,width=12cm]{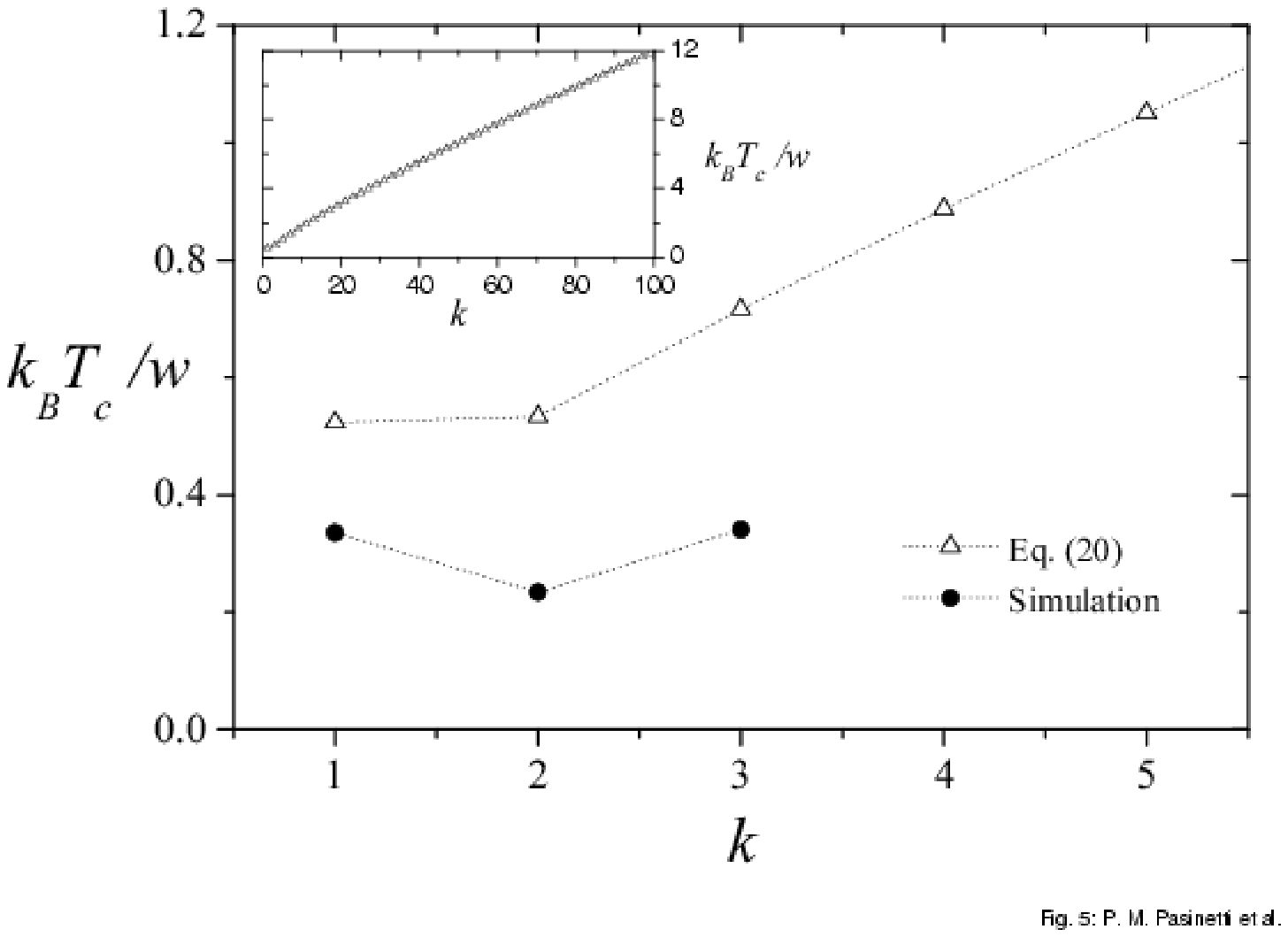}
\end{figure}
\begin{figure}
\includegraphics[angle=0,width=12cm]{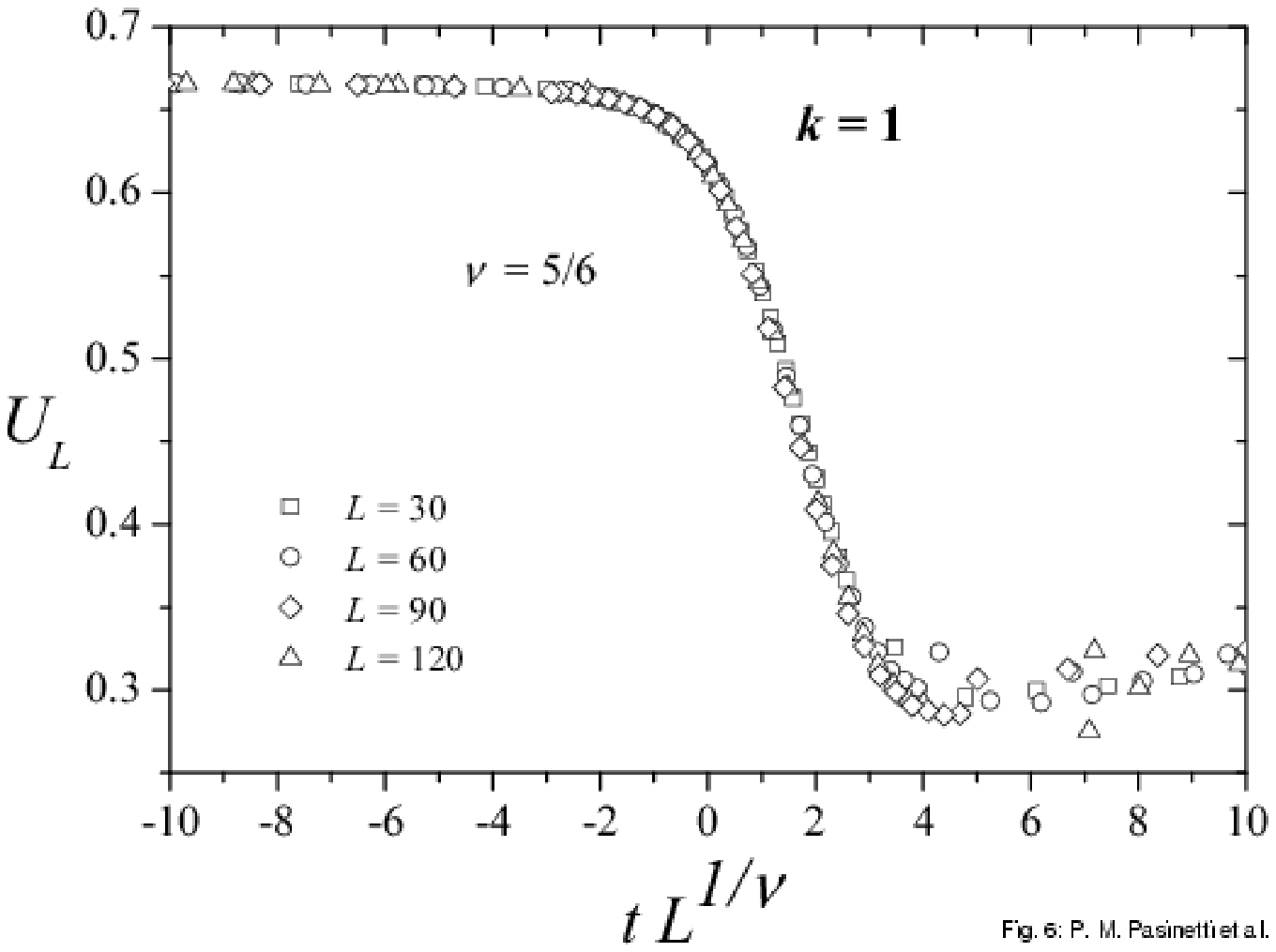}
\end{figure}
\begin{figure}
\includegraphics[angle=0,width=12cm]{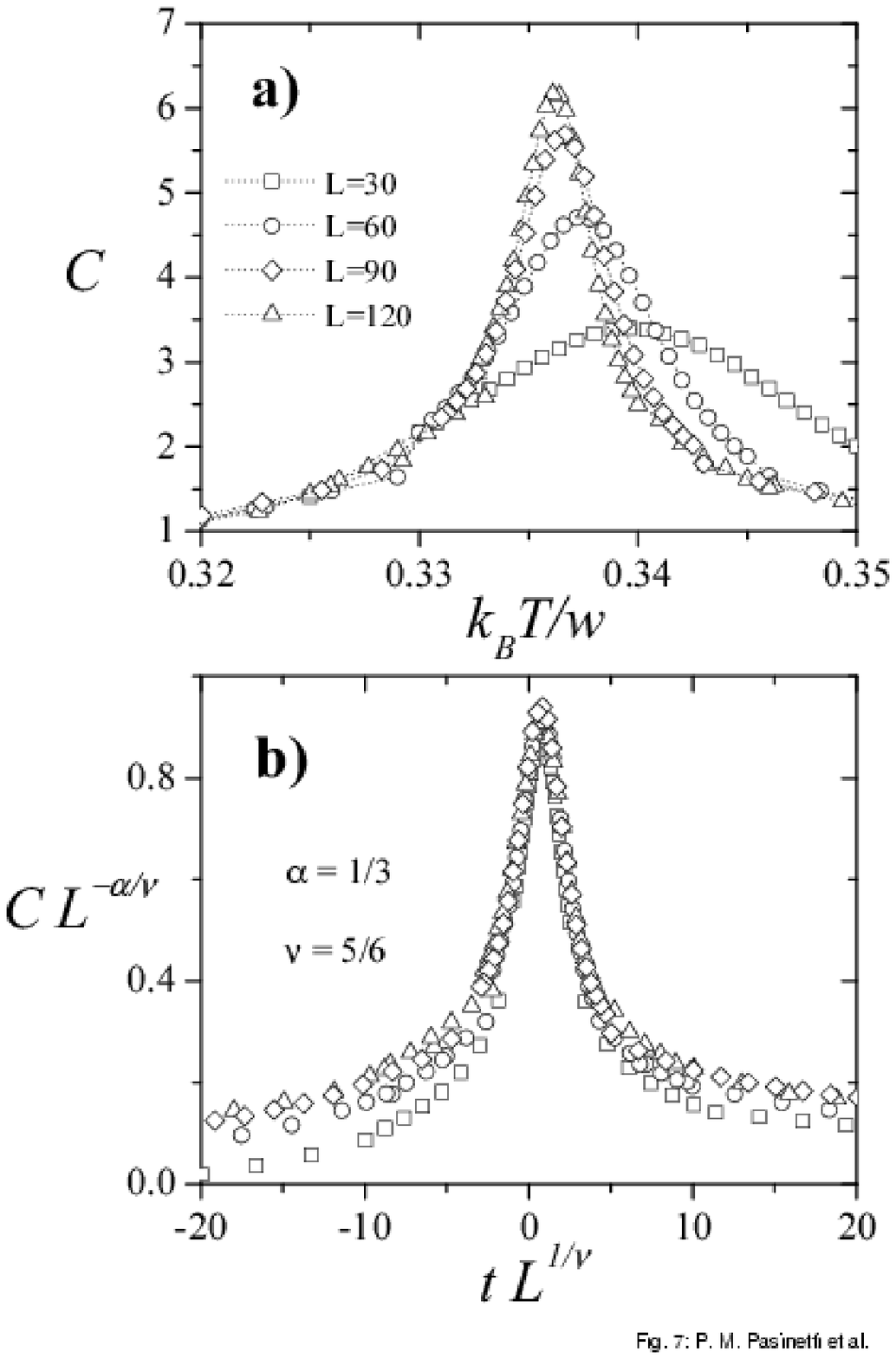}
\end{figure}
\begin{figure}
\includegraphics[angle=0,width=12cm]{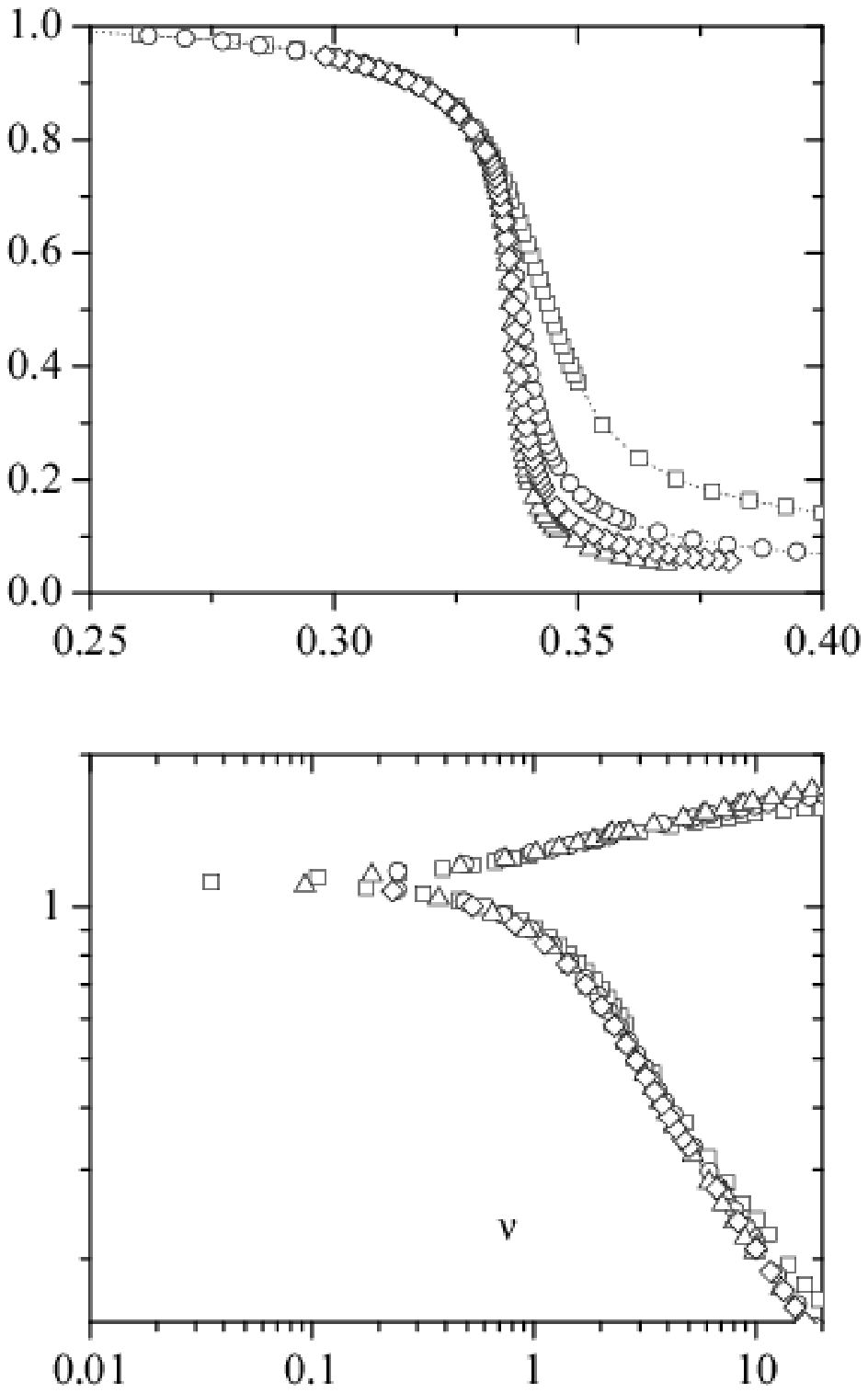}
\end{figure}
\begin{figure}
\includegraphics[angle=0,width=12cm]{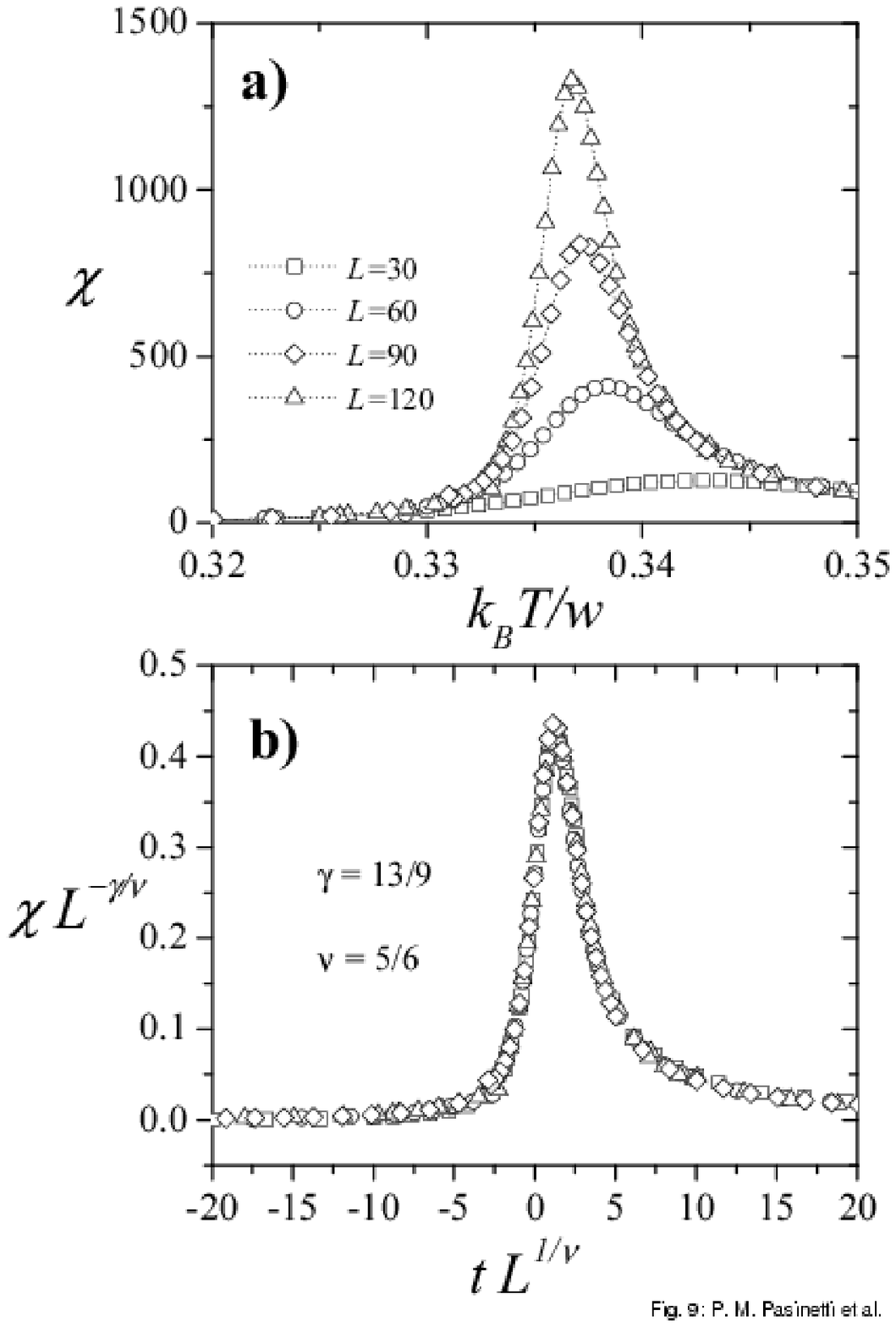}
\end{figure}
\begin{figure}
\includegraphics[angle=0,width=12cm]{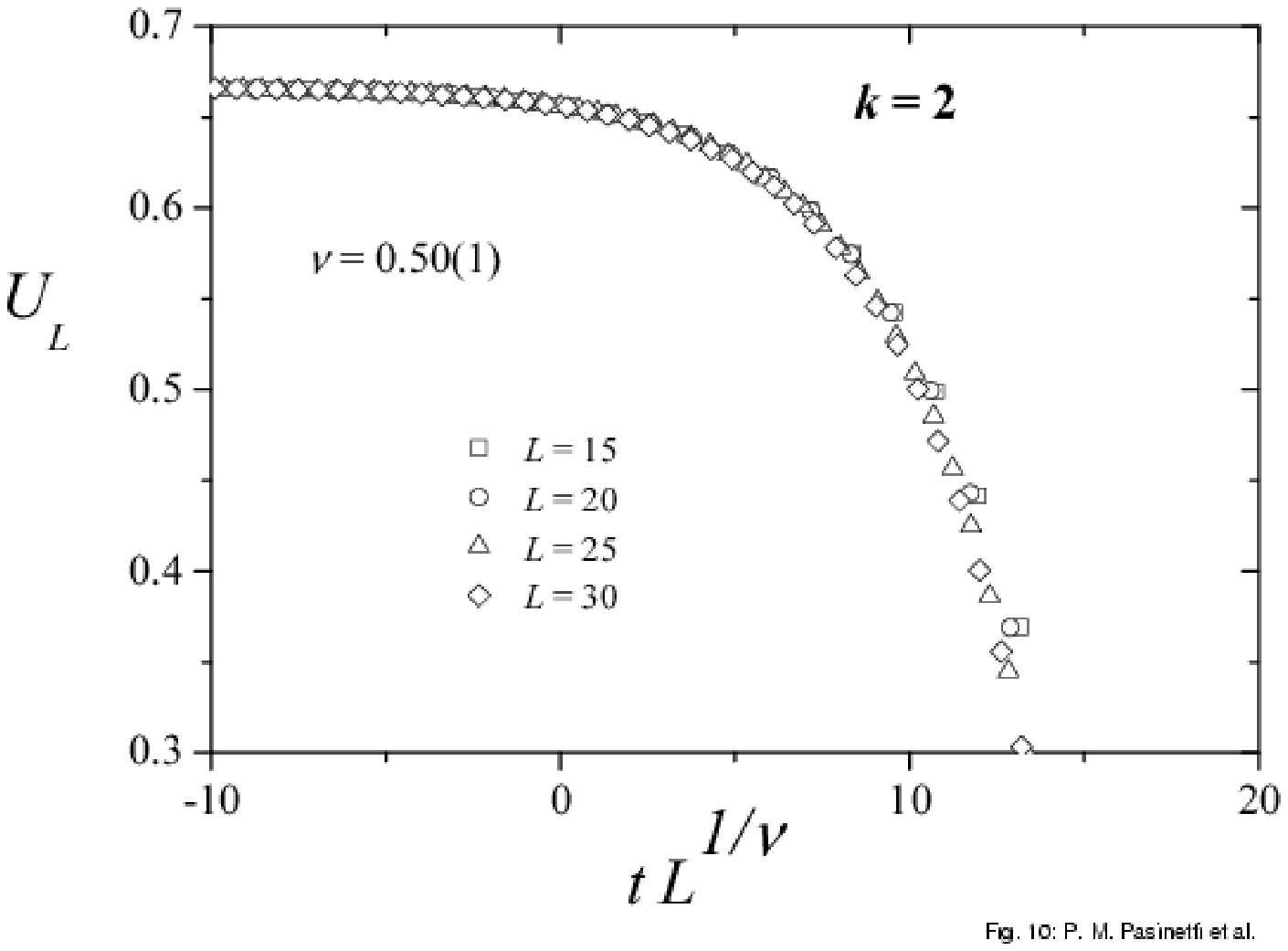}
\end{figure}
\begin{figure}
\includegraphics[angle=0,width=12cm]{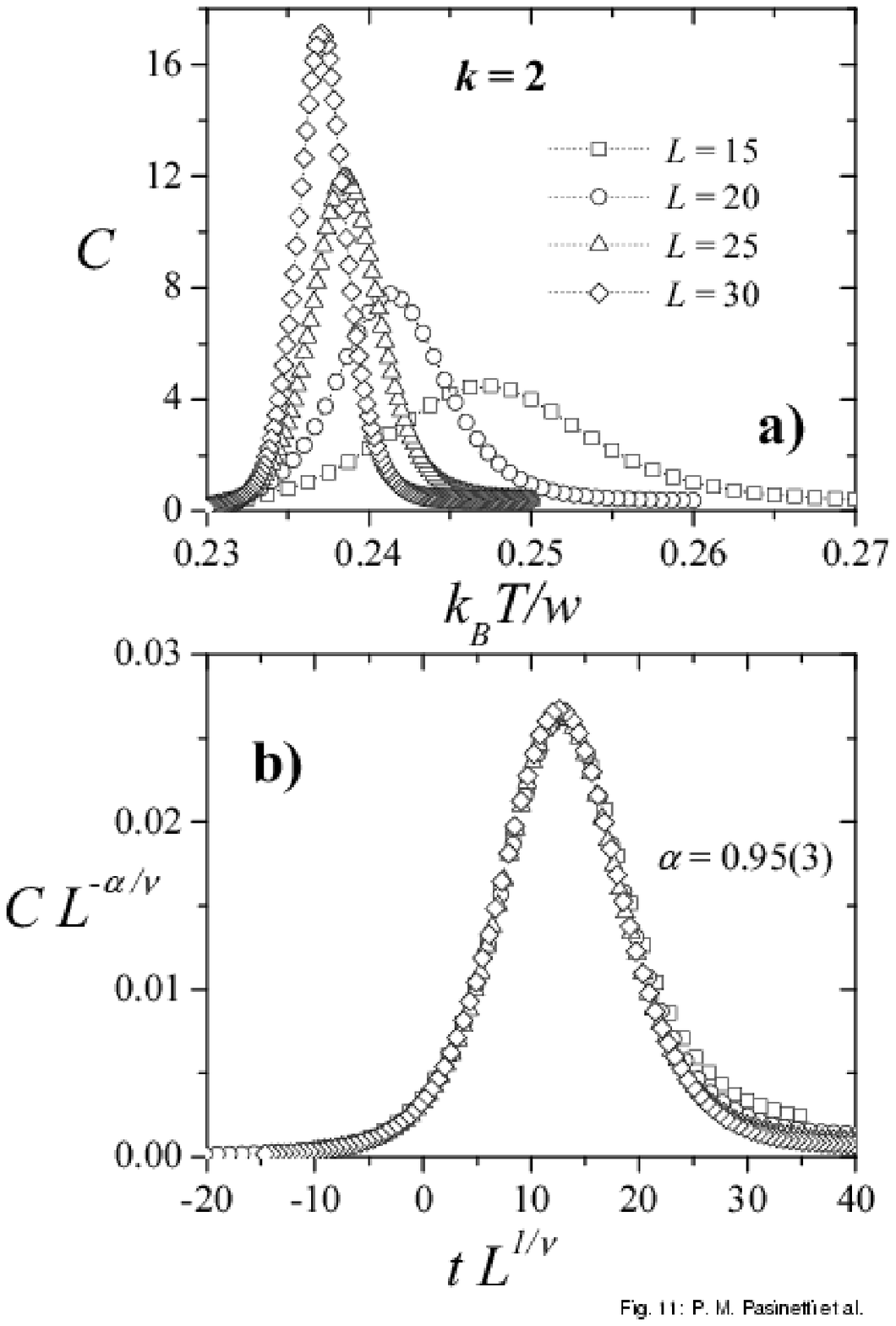}
\end{figure}
\begin{figure}
\includegraphics[angle=0,width=12cm]{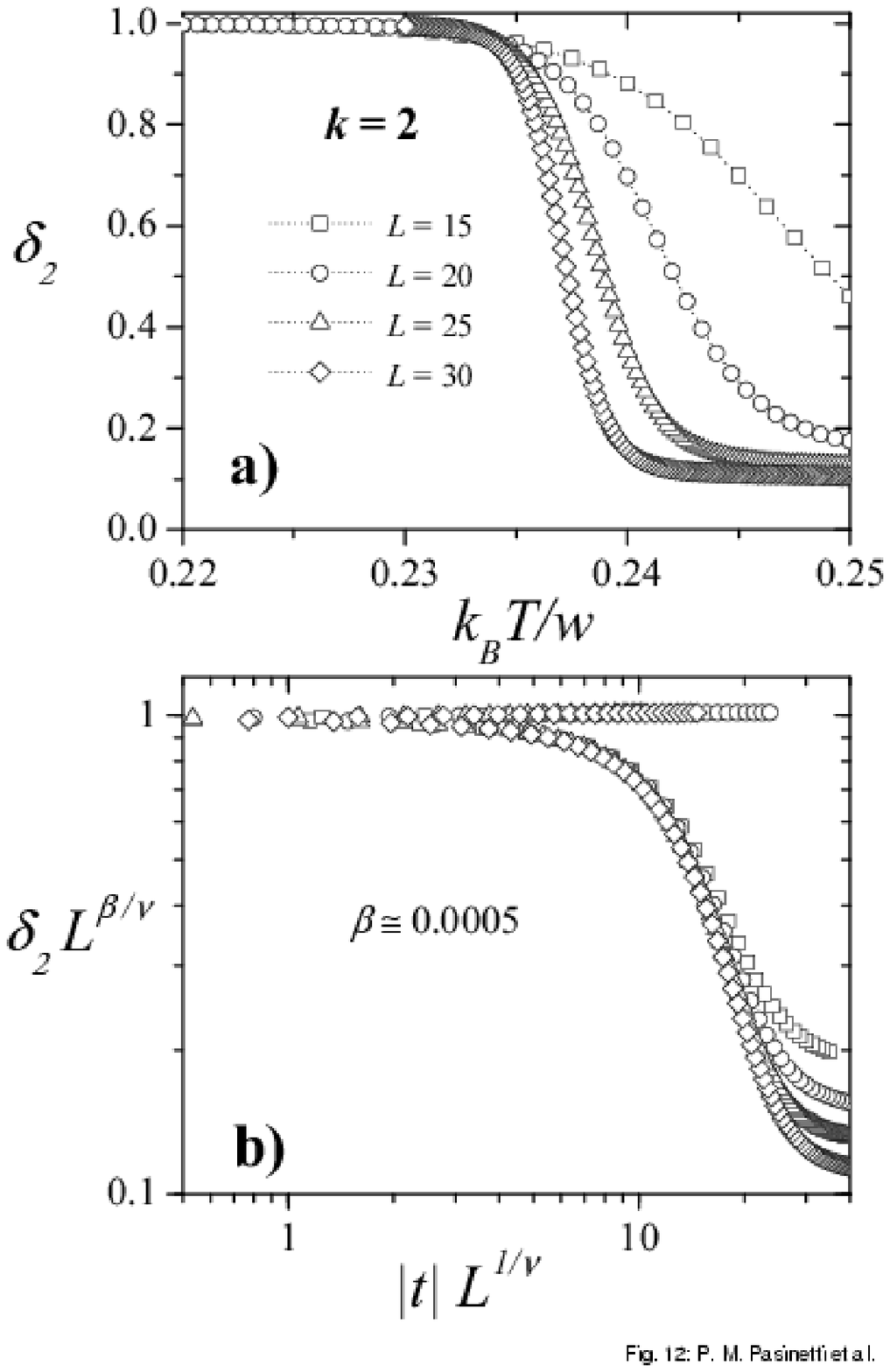}
\end{figure}
\begin{figure}
\includegraphics[angle=0,width=12cm]{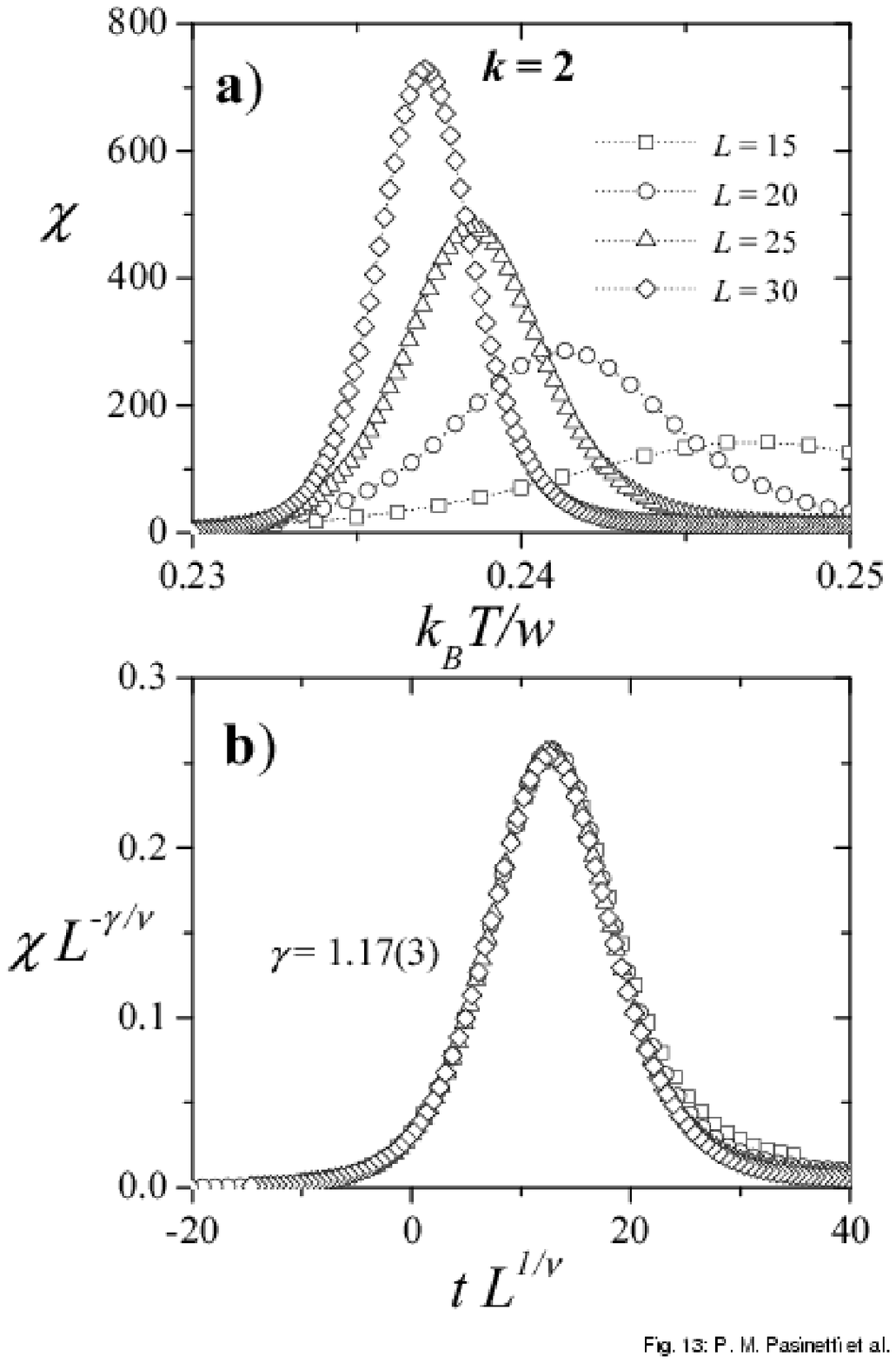}
\end{figure}


\begin{thebibliography}{10}
\bibitem{foot1} The diffusional relaxation is applied only for
dimers and trimers.
\bibitem{foot2} Even though this criterion has been widely used in the
literature in order to identify the universality class of a given
transition (see for example Ref.\cite{Rzysko}), it is not easy to
establish a theoretical framework of general applicability. In
fact, recent studies (on models with the same universality class)
have found a non-universal behavior of $U^*$ \cite{Selke}.
\end{thebibliography}
\end{document}